\documentclass[prl,aps,showpacs,twocolumn]{revtex4-1}
\usepackage{psfrag}
\usepackage{graphicx}
\usepackage{dcolumn}
\usepackage{color}
\usepackage{latexsym,amsfonts}
\usepackage{bm}
\usepackage{amssymb}
\begin{document}

\title{Parity--Even and Time--Reversal-Odd Neutron Optical Potential
  in Spinning Matter Induced by Gravitational Torsion}
\author{A. N. Ivanov} \email{ivanov@kph.tuwien.ac.at}
\affiliation{Atominstitut, Technische Universit\"at Wien, Stadionallee
  2, A-1020 Wien, Austria}
\author{W. M.~Snow}\email{wsnow@indiana.edu} \affiliation{Indiana
  University, Bloomington, Indiana 47408, USA} \affiliation{Center for
  Exploration of Energy and Matter, Indiana University, Bloomington,
  IN 47408} \date{\today}
\begin{abstract}

Recent theoretical work has shown that spin $1/2$ particles moving
through unpolarized matter which sources torsion fields experience a
new type of parity-even and time-reversal-odd optical potential if the
matter is spinning in the lab frame. This new type of optical
potential can be sought experimentally using the helicity dependence
of the total cross sections for longitudinally polarized neutrons
moving through a rotating cylindrical target. In combination with
recent experimental constraints on short-range P--odd, T--even torsion
interactions derived from polarized neutron spin rotation in matter
one can derive separate constraints on the time components of scalar
and pseudoscalar torsion fields in matter. We estimate the sensitivity
achievable in such an experiment and briefly outline some of the
potential sources of systematic error to be considered in any future
experimental search for this effect.

\end{abstract}
\pacs{13.88.+e, 13.75.Cs, 14.20.Dh, 14.70.Pw}
\maketitle
\section{Introduction}

Ever since Einstein's theory of general relativity (GR) successfully
proposed an intimate connection between the geometry of spacetime and
its matter content, physicists have been encouraged to consider the
geometric structure of spacetime as a legitimate subject for
scientific study. Among the mathematical quantities that characterize
such geometries are curvature and torsion. GR makes essential use of
curvature: gravity is interpreted as spacetime curvature and test
particle trajectories are geodesics. Spacetime torsion is the other
natural geometric quantity that is available to characterize spacetime
geometry. Although torsion vanishes in GR, many models which extend GR
include various types of nonvanishing torsion sourced by some form of
spin density~\cite{torsion}. Yet experimental searches for
gravitational torsion are usually specific to a particular torsion
model. Even if the coupling of torsion to spin is similar in strength
to that of curvature to the energy--momentum tensor, strong
spin-density sources which could generate observable effects are
difficult to realize. A large fraction of the previous work on
gravitational torsion is theory-centric and attempts to argue for
specific realizations of torsion in particular theories coming from
various mathematical and physical motivations.

By contrast in this work we treat the question of the presence of
torsion as an issue to be answered by experiment and make no
theoretical assumptions about its possible strength or range. This
intellectual perspective favors a qualitatively different experimental
strategy which can catch many different torsion possibilities at
once. Torsion interactions which violate discrete symmetries can be
sought with high sensitivity and can benefit from the many powerful
techniques of precision measurement which have been developed to
search for discrete symmetry violation outside of gravitational
physics. We therefore believe that the new possibilities for
experimental investigation of torsion along the lines discussed in
this Letter are of general interest in the physics community.

Tight model-independent constraints on the size of a very broad set of
long-range torsion background fields in spacetime have already been
derived from the intellectual perspective we advocate through the
appropriate reinterpretation of experiments designed to search for
Lorentz and CPT violation~\cite{CL97,krt08,Obukhov2014}. This work
derived stringent constraints on 19 of the 24 components of a possible
ambient torsion field ${\cal T}^{\alpha}{}_{\mu\nu}(x)$ through the
coupling of components $T^\mu$, $A^\mu$ and ${\cal
  M}^{\alpha}{}_{\mu\nu}$ of its irreducible representation
\cite{krt08} to fermions in a general effective Lagrange density with
all independent constant-torsion couplings of mass dimensions four and
five. Torsion fields which do not extend far from sources would not be
seen in the experiments used in these analyses. It is therefore of
interest to consider how one might constrain a broad set of
possible short-range torsion fields experimentally.

We argue that the most promising experimental observable for the type
of broadband torsion searches that we advocate are coherent
spin-dependent optical effects in forward scattering. No matter what
the range of the torsion fields sourced by fermionic matter, such
fields must contribute to the forward scattering amplitude of a spin
$1/2$ particle by the optical theorem of scattering theory. Given the
form of any particular torsion model one could easily evaluate its
contribution to the forward amplitude and therefore make direct
contact with experimental bounds. Coherent spin-dependent effects in
forward scattering can be sought experimentally with high sensitivity
using quantum interference. Torsion interactions which violate
discrete symmetries are best to look for as they are relatively
insensitive to background effects from other physical processes.
Polarized slow neutrons in particular are an excellent choice for such
an experimental investigation. Neutrons constitute a massive spin
$1/2$ probe which can penetrate macroscopic amounts of matter due to
their zero electric charge and lack of ionizing interactions with
matter, and they can also be used to perform sensitive polarization
measurements using various types of interferometric methods.

We therefore focus our attention on polarized neutron optical effects
induced by torsion interactions which violate parity and time reversal
symmetry in P--odd/T--even, P--even/T--odd, and P--odd/T--odd
combinations. Recently the first experimental upper bound has been
set~\cite{Lehnert2015} on the optical potential from P--odd and
T--even short-range torsion fields. The experiment employed
transversely polarized slow neutrons that traversed a meter of liquid
$^{4}$He. Torsion fields sourced by the protons, neutrons, and
electrons in the helium atoms can generate a term in the slow-neutron
optical potential proportional to $\vec{\sigma} \cdot \vec{p}$. The
$\vec{\sigma}\cdot\vec{p}$ term in the neutron optical potential
violates parity and therefore causes a rotation of the plane of
polarization of a transversely polarized slow neutron beam about its
momentum as it moves through matter~\cite{Mic64, Forte80, Heckel82,
  Heckel84}. The rate of rotation of the neutron's spin about
$\vec{p}$ may be characterized by the neutron rotary power $d\phi_{\rm
  PV}/dL$, where $\phi_{\rm PV}$ denotes the angle of rotation and $L$
the distance the neutron has traversed in the sample. For the Lagrange
density above in the nonrelativistic limit, $d\phi_{\rm PV}/dL =
2\zeta $ where $\zeta$ is a liner combination of the scalar $T_0$ and
pseudoscalar $A_0$ torsion components equal to $\zeta = (2m
\xi^{(5)}_8 - \xi^{(4)}_2)\,T_0 + (2m \xi^{(5)}_9 -\xi^{(4)}_4)\,A_0$,
where $m$ is the neutron mass and $\xi^{(5)}_8$, $\xi^{(4)}_2$,
$\xi^{(5)}_9$ and $\xi^{(4)}_4$ are phenomenological constants defined
in~\cite{krt08,Lehnert2015}.  The limit on $\zeta$ from this work was
$|\zeta| < 9.1 \times 10^{-23}$ GeV. Later work~\cite{Ivanov2015a}
showed that the limit on $\zeta$ from long-range torsion fields using
other data could be further improved by 5 orders of magnitude. This
measurement constrains a linear combination of possible internal
torsion fields of arbitrary range generated by the spin-$\frac{1}{2}$
protons, neutrons, and electrons in the helium. Although future
neutron spin rotation experiments could in principle be used to set
more stringent torsion constraints, in practice measurements of this
type if pushed to higher precision will encounter a background
parity-odd spin rotation from the neutron-nucleus weak interaction in
the Standard Model. Although this background is calculable in
principle, in practice our inability to perform calculations involving
the strong interaction for low energy processes makes it impractical
to subtract off the Standard Model contribution to parity violation in
this case. We therefore do not anticipate further significant
experimental improvements on P--odd neutron-torsion interactions from
measurements of this type.


It is interesting to ask whether or not there are other experimental possibilities using slow neutrons which can access short-range torsion effects in matter. In this Letter we point out that the answer to this question is yes if one analyzes neutron optical effects in nonstationary media. The existence of such a term has been demonstrated recently by Ivanov and Wellenzohn~\cite{Ivanov2016}. They show that a nonrelativistic spin $1/2$ particle moving in a medium rotating with angular velocity $\vec{\omega}$ in the presence of a scalar neutron-torsion coupling can possess a P--even and T--odd term in the neutron-matter optical potential of the form 
\begin{eqnarray}
\label{eq:1a} 
\Phi^{(\rm T-odd)}_{\rm eff} = - \frac{2}{3}\,\frac{{\cal E}_0}{m}\,i\, \vec{\sigma}\cdot \vec{\omega}, 
\end{eqnarray} 

where ${\cal E}_0$ is the scalar component of the torsion field, which
is equal to ${\cal E}_0 = - T_0$ in notations of
Kosteleck\'y~\cite{krt08,Kostelecky2004}, and $\vec{\omega}$ is the
angular velocity of the cylinder rotating around the $z$--axis of the
direction of motion of the neutron beam. Below for closer connection
to the notation used by Lehnert {\it et al.} \cite{Lehnert2015} we set
${\cal E}_0 = - T_0$. Note that a measurement of this effect in
comparison with the existing data from neutron spin rotation can
separate the scalar $T_0$ torsion component from the pseudoscalar
$A_0$ torsion component. The appearance of this P--even and T--odd
torsion--Dirac fermion potential has a geometrical
origin~\cite{Obukhov2009, Ivanov2016}. Hadley~\cite{Hadley2011}
identified the scalar field equal to the frame dragging term $d\phi
\over dt$ in the Kerr metric of a spinning massive body as a source
for violation of CP--invariance, which is related to violation of
T--invariance assuming CPT conservation. In contrast to the P--odd
torsion--neutron interaction proportional to $\vec{\sigma}\cdot
\vec{p}$ discussed earlier, this P--even and T--odd torsion--neutron
potential Eq.(\ref{eq:1a}) is proportional to $\vec{\sigma}\cdot
\vec{\omega}$. The contribution of the potential Eq.(\ref{eq:1a}) to
the forward amplitude in low—energy neutron--nucleus scattering for
neutrons of momentum $p$ is given by
\begin{eqnarray}\label{eq:2a}
f_{\rm TV}(0) = - i\,\frac{1}{3}\,T_0 R^2 L\,\omega\,
\varphi^{\dagger}_{\rm out}\sigma_z\varphi_{\rm in},
\end{eqnarray}
where $R$ and $L$ are the radius and length of a right circular
cylinder rotating around the $z$--axis. This dependence of the forward
amplitude Eq.(\ref{eq:2a}) on the parameters of a rotating cylinder is
caused by the existence of the effective T--odd potential
Eq.(\ref{eq:1a}) inside the cylinder~\cite{Ivanov2016}. $\varphi_{\rm
  in}$ and $\varphi_{\rm out}$ are the column Pauli spinors of the
neutron in the initial and final state, respectively. They are
eigenfunctions of the operator $\vec{\sigma}\cdot \vec{n}$,
i.e. $(\vec{\sigma}\cdot \vec{n}\,)\varphi = \pm\,\varphi$, where
$\vec{n}$ is a unit vector of the neutron position inside the rotating
cylinder, characterized by the polar $\theta$ and azimuthal $\phi$
angles, and $\pm 1$ are the neutron spin polarizations. Assuming that
in the initial state neutrons are polarized parallel and
anti--parallel the $x$--axis with the wave functions
$\varphi^{(\pm)}_{\rm in}$ having the following elements $(\pm
1/\sqrt{2}, 1/\sqrt{2})$ and in the final state neutrons are described
by the wave functions $\varphi^{(\pm)}_{\rm out}$ with elements
$(\cos(\theta/2), \sin(\theta/2)\,e^{\,-i\phi})$ and $(-
\sin(\theta/2)\,e^{\,+i\phi},\cos(\theta/2))$, respectively, the
T--odd contributions to the s--wave amplitude of scattering of
polarized neutrons by nucleus are given by
\begin{eqnarray}\label{eq:3a}
f^{(\pm)}_{\rm TV}(0) = \mp i\,\frac{T_0}{3\sqrt{2}}\,R^2 L\,\omega\,\Big( \cos\frac{\theta}{2} -\sin\frac{\theta}{2}\, e^{\,\pm\,i\,\phi}\Big).
\end{eqnarray}
According to Stodolsky~\cite{Sto82}, the contribution of T--odd
interaction to the cross section of low energy neutron--nucleus
scattering is given by $\sigma_{\rm TV} = (4\pi/p)\,{\rm Im}\Delta
f_{\rm TV}(0)$, where $p$ is a neutron momentum and $\Delta f_{\rm
  TV}(0) = f^{(+)}_{\rm TV}(0) - f^{(-)}_{\rm TV}(0)$. Using
Eq.(\ref{eq:3a}) for the T--odd contribution to the cross section we
obtain the following expression
\begin{eqnarray}\label{eq:4a}
\Delta \sigma_{\rm TV} = -\frac{8\pi}{3 \sqrt{2}}\,T_0\,R^2 L\,
\frac{\omega}{p}\,\Big( \cos\frac{\theta}{2} - \sin\frac{\theta}{2}\,
\cos\phi\Big).
\end{eqnarray}
To avoid certain systematic effects which can be induced by a spinning
cylinder \cite{Barnett1915,Barnett1935}, the neutrons should be
polarized parallel and anti--parallel the $z$--axis with the wave
functions, which can be obtained from the wave functions
$\varphi^{(\pm)}_{\rm out}$ at $\theta = 0$. Setting $\theta = 0$ we
get
\begin{eqnarray}\label{eq:5a}
\Delta \sigma_{\rm TV} = - \frac{8\pi}{3 \sqrt{2}}\,T_0\,R^2
L\,\frac{\omega}{p}.
\end{eqnarray}
The experiment would then search for the $\omega$--dependent part of
the helicity-dependent component of the polarized neutron cross
section difference for neutrons passing through a cylinder rotating
with an angular velocity $\omega$ and be sensitive to the scalar
torsion parameter $T_0$.

At first glance the $1/p$ dependence in $\Delta \sigma_{TV}$ may look
strange, since a well--known result from nonrelativistic scattering
theory shows that the imaginary part of the forward elastic scattering
amplitude tends to zero in the limit $p R<<
1$~\cite{Landau1958}. However this argument does not directly apply to
our case of a T--odd torsion optical potential inside matter. A purely
imaginary term in the forward scattering amplitude proportional to
$1/p$ for the T--odd component of $\Delta \sigma_{TV}$ is fully
consistent with unitarity. At some point in the extreme $p \to 0$
limit the finite size of the extent of the medium, if nothing else,
will eventually come into play and give a finite contribution to the
cross section which will prevent $\Delta \sigma_{TV}$ from diverging.

The P--even and T--odd nature of this observable is quite insensitive
to potential backgrounds from known interactions. P--even and T--odd
interactions involving Standard Model fields require a violation of C
which can be introduced neither at the first generation quark level
nor into the gluon self- interaction. Consequently, one needs to
consider C violation between quarks of different generations and/or
between interacting fields. P--even and T--odd interactions between
identical spin $1/2$ fermions vanish
identically~\cite{Khriplovich1991}. Indirect constraints from analyses
of radiative corrections to constraints on P--odd and T--odd
interaction from electric dipole moment searches~\cite{Haxton1994} are
more stringent than the direct experimental constraints. No P--even
and T--odd physical effect has ever been observed experimentally. The
most sensitive direct experimental upper bounds on P--even and T--odd
interactions of the neutron come from an analysis~\cite{Simonius1997}
of measurements of charge symmetry breaking in neutron-proton elastic
scattering~\cite{Abegg1989,Vigdor1992, Zhao1998} and a
polarized-neutron transmission-asymmetry experiment using transversely
polarized $5.9\,$MeV neutrons in a nuclear spin-aligned target of
holmium~\cite{Huffman1997}. Sensitive experiments to search for
P--even and T--odd angular correlations in neutron beta
decay~\cite{Kozela2009, Mumm2011} have seen no such effects. The
observable considered in this work is therefore especially insensitive
to possible contamination from other physical effects.

Now we briefly discuss the potential sources of systematic error which
might be involved in a measurement of this P--even and T--odd term in
the forward scattering amplitude from torsion interactions in the
presence of spinning matter proportional to $\vec{\sigma} \cdot
\vec{\omega}$. To our knowledge such a measurement has not been
considered in the literature. The most worrisome potential systematic
effect would be a physical phenomenon which makes an internal magnetic
field or a spin polarization in the medium proportional to
$\omega$. Such a physical phenomenon exists and is known as the
Barnett effect~\cite{Barnett1915, Barnett1935}, the time-reversed
version of the more well-known Einstein-deHass-Alfven
effect~\cite{Einstein1915}. In a rotating medium with a finite
magnetic susceptibility, the orbital and spin angular momentum vectors
which are responsible for the magnetic susceptibility of
non-ferromagnetic media will tend to align with $\vec{\omega}$ and
will produce a magnetization in the medium $B={\omega \over \gamma}$
where $\gamma$ is the gyromagnetic ratio of the sample. The Barnett
effect was observed long ago in ferromagnetic media and has recently
been observed experimentally for the first time in a paramagnetic
spinning medium~\cite{Ono2015} in gadolinium, which possesses a very
large magnetic susceptibility $\chi$ and an internal magnetization
$M=\chi B$ of $30 nT$ for $\omega=10^{4}$ Hz. This effect can be
greatly suppressed by using a material with a low magnetic
susceptibility. In addition, any unpaired electrons or nucleons in
such a rotating medium thereby get polarized and can interact with the
polarized neutrons through either the electromagnetic interaction or
the spin-dependent strong interaction to generate a spin-dependent
term in the total cross section proportional to $\vec{\sigma} \cdot
\vec{I}$ where $\vec{I}$ is the relevant rotation-induced nuclear or
electron polarization~\cite{Bar12, Bar65, Abr72}. However both of
these effects are T--even and therefore the corresponding forward
amplitudes are out of phase by $\pi/2$ with respect to the T--odd
effect considered in this work.

A wide variety of possible neutron spin rotation effects in
noncentrosymmetric structures that could be induced by
rotationally-generated stresses in matter have been estimated
theoretically for slow neutrons~\cite{Kabir1974, Kabir1975, Karl1976,
  Cox1977a, Cox1977b, Ritchie1979} and are very small, since the
effects of the chiral electronic structure must be dynamically
communicated somehow to the nuclear motion, and often this can only be
done through higher-order electromagnetic effects. The passage of slow
neutrons through an accelerating material medium produces energy
changes in the neutron beam if the boundaries are accelerating
according to arguments using the equivalence
principle~\cite{Kowalski1993, Kowalski1995, Littrell1996, Nosov1998}
and have been recently resolved experimentally using measurements with
ultracold neutrons~\cite{Frank2006, Frank2008, Frank2011, Frank2012}
but vanish for the case of interest in this work.  Various effects
involving rotating neutron optical elements~\cite{Mashhoon1998,
  Mashhoon2006, Mashhoon2009} also do not generate our effect.

Using obvious choices for the material of the spinning cylinder
($MgF_{2}$, silicon) which possess a long neutron mean free path with
minimal neutron absorption and small angle scattering and are composed
of light nuclei which do not possess low-lying neutron-nucleus
resonances, one could achieve a sensitivity to $T_{0}$ of $10^{-32}$
GeV in a practical experiment. Comparing with the existing
constraints~\cite{Lehnert2015} and \cite{Ivanov2015a} on the linear
combination of $T_{0}$ and $A_{0}$ described above, this is of about
10 and 5 orders of magnitude smaller.

The other obvious choice which meets our criteria, namely a P--odd and
T--odd torsion-dependent term in the neutron forward scattering
amplitude, is also possible in principle~\cite{Bar12}. A P-odd and
T-odd term in the forward scattering amplitude can be accessed
experimentally in polarized neutron optics if the target medium is
also polarized. Such an observable can indeed access types of
gravitational torsion interactions distinct from the ones discussed
above~\cite{Bonder2016}. However the large spin dependence of the
neutron-nucleus strong interaction would create severe difficulties
for experimental torsion searches of this type. One could realize such
a search in practice for neutron-electron torsion couplings by
employing special materials which possess nonzero electron
polarization and small internal magnetic fields~\cite{Leslie2014}.

We have pointed out that the recently-identified P--even and T--odd
effects induced by effective low--energy torsion--neutron interactions
in rotating media can be sought experimentally by measuring the
helicity dependence of the total cross section for neutrons moving
through a spinning cylinder. The difference of the cross sections of
oppositely polarized neutrons caused by the effective low--energy
P--even and T--odd potential Eq.(\ref{eq:1a}), depends linearly on an
angular velocity of a rotating cylinder. Such an experiment can access
the time component of short-range torsion fields sourced by the atoms
in the medium and is sensitive to a different set of torsion fields
compared to previous experimental work sensitive to P-odd short-range
torsion. We considered a number of potential sources of systematic
error in an experiment of this type. We are encouraged that with
careful design such an experiment can be conducted with negligible
systematic error. Finally we would like to note that, according to a
recent analysis of cosmological constant or dark energy density as
induced by torsion fields~\cite{Ivanov2016a}, the measurements of
torsion in terrestrial laboratories could shed light on the origin of
the Universe creation and dark energy as a relic of the Universe
evolution.

\subsection{Acknowledgements}

The work of A. N. Ivanov was supported by the Austrian ``Fonds zur
F\"orderung der Wissenschaftlichen Forschung'' (FWF) under contracts
1689-N16, I862-N20 and P26781-N20. The work of W. M. Snow was
supported by US National Science Foundation grant PHY-1306942, by the
Indiana University Center for Spacetime Symmetries, and by the Indiana
University Collaborative Research and Creative Activity Fund of the
Office of the Vice President for Research. W. M. Snow also
acknowledges discussions with A. Kosteleck\'y and Y. Bonder on the
general subject of gravitational torsion theory and with B. Mashhoon
on the dynamics of polarized neutrons in spinning matter.

\end{document}